\documentstyle[preprint,eqsecnum,aps]{revtex}
\begin{document}
\draft
\title{Quantum Conditions of the Relativistic Potential Problems }
\author{De-Hone Lin\thanks{%
e-mail: d793314@phys.nthu.edu.tw}}
\address{Department of Physics, National Tsing Hua University \\
Hsinchu 30043, Taiwan}
\date{\today}
\maketitle

\begin{abstract}
The quantum conditions of the relativistic integrable systems whose
classical motion is multiply periodic are given by considering the
single-valuedness of the linear superposition of the approximate solutions $%
R_{i}\exp {\{iS_{i}/\hbar \}}$ with $R_{i}$ and $S_{i}$ being the solutions
of the different branches of the current conservation equation and the
relativistic Hamilton-Jacobian equation, respectively.
\end{abstract}
\newpage \tolerance=10000
\section{Introduction}

~~~~~~After Planck's and Einstein's works on the electromagnetic radiation,
it was Niels Bohr who first proposed a feasible theory for the energy
spectrum of the hydrogen atom. He pointed out that the motion of a
microscopy system can be described by classical mechanics via restricting
certain constants of the motion to be integers. These constraint conditions
are called quantum conditions. The integers arising in them are called
quantum numbers.

In 1917, Einstein \cite{1} showed that the Bohr quantum conditions could be
generalized in an invariant manner because of $\sum p_{i}dq_{i}$ being
invariant via a contact transformation of variable to include systems whose
classical motion is not separable in the coordinates provided that the
classical motion is multiply periodic. Another derivation of the conditions
by way of the Schr\"{o}dinger equation was given by Brillouin \cite{2}. He
explained by considering the phase of approximate solution single value how
Einstein's conditions was a consequence of the single-valuedness of the wave
functions of the quantum mechanics. Nevertheless, there is no principle can
be used to explain how to determined the integer or half integer for quantum
numbers in any particular cases.

For separable systems, a technique of deciding the integer or half integer
quantum numbers was found by Kramers \cite{3} by means of an approximate
solution of Schr\"{o}dinger equation. He also showed that these conditions
are consequences of quantum mechanics in the limit as Planck's constant $h$
tends to zero. Furthermore, Keller \cite{4} pointed out that by considering
the multivaluedness of phases and amplitudes of the approximate solutions
the conjugate points in the classical motion require a slight modification
of Brillouin's and Einstein's conditions.

In this paper we follow the general recipe proposed by Keller in Ref. \cite
{4} where the quantum conditions of the non-relativistic potential problems
was discussed. We give the quantum conditions of the relativistic integrable
system by assuming the single-valuedness of the approximate solution $%
\varphi ({\bf x})$ of Klein-Gordon equation whose amplitudes and phases of
the components are given by the relativistic current conservation equation
and Hamilton-Jacobian equation, respectively. Not only for the more
completeness of the theories is this but for the more accurate description
of the relativistic semiclassical physics. The reason is that in the
semiclassical relativistic physics a correct quantum condition for the
relativistic systems is important in determining the density of states, wave
functions etc. which provide us the informations of the relativistic quantum
mechanics in terms of the quantities of the relativistic classical mechanics 
\cite{lk}.

\section{Quantum conditions for the relativistic integrable systems}

With the metric $g_{\mu \nu }={\rm diag}(1,-1,-1,-1)$, the electromagnetic
field is described by the four-vector 
\[
A^{\mu }=\{A_{0},{\bf A\}=\{}A_{0},A_{x},A_{y},A_{z}\}=g^{\mu \nu }A_{\nu } 
\]

\[
A_{\mu }=g_{\mu \nu }A^{\nu }=\{A_{0},-{\bf A\}} 
\]

In the case of non-relativistic quantum mechanics we specified minimal
coupling of the electromagnetic field, 
\begin{equation}
\hat{E}\Rightarrow i\hbar \frac{\partial }{\partial t}-eA_{0}\quad ,\quad 
\hat{p}\Rightarrow -i\hbar \nabla -\frac{e}{c}{\bf A,}  \label{1}
\end{equation}
which can be compressed to the four-dimensional and covariant form as 
\begin{equation}
\hat{p}^{\mu }\Rightarrow \hat{p}^{\mu }-\frac{e}{c}A^{\mu }\quad ,\quad 
\hat{p}_{\mu }\Rightarrow \hat{p}_{\mu }-\frac{e}{c}A_{\mu }.  \label{2}
\end{equation}
With the same minimal coupling, the Klein-Gordon equation with an
electromagnetic field is given by 
\begin{equation}
\left( \hat{p}^{\mu }-\frac{e}{c}A^{\mu }\right) \left( \hat{p}_{\mu }-\frac{%
e}{c}A_{\mu }\right) \psi =m^{2}c^{2}\psi  \label{3}
\end{equation}
or explicitly 
\begin{equation}
\frac{1}{c^{2}}\left( i\hbar \frac{\partial }{\partial t}-eA_{0}\right)
^{2}\psi =\left( \left( i\hbar \nabla +\frac{e}{c}{\bf A}\right)
^{2}+m^{2}c^{2}\right) \psi .  \label{4}
\end{equation}
In derivation of the paper, we let simply $eA_{0}=V({\bf x})$ and ${\bf A=}0$%
. The vector fields can be simply considered by taking $p\rightarrow p-e/c%
{\bf A}$ in the final result. Then the stationary Klein-Gordon equation
reads, letting the time dependent wave function $\psi ({\bf x},t)=\varphi (%
{\bf x})e^{-iEt/\hbar }$, 
\begin{equation}
\left[ \left( \hbar c\right) ^{2}\nabla ^{2}+\left( E-V({\bf x})\right)
^{2}-m^{2}c^{4}\right] \varphi ({\bf x})=0.  \label{5}
\end{equation}
To derive the quantum condition, let us assume that the $\varphi ({\bf x})$
is approximately equal to 
\begin{equation}
\varphi ({\bf x})=\sum_{i}R_{i}e^{\frac{i}{\hbar }S_{i}},  \label{6}
\end{equation}
where the assumption of the linear combination property of the wave
functions arise in almost all microscopic problems. It stands for the wave
characteristic of microscopic scale. On inserting the expression Eq. (\ref{6}%
) for $\varphi ({\bf x})$ into Eq. (\ref{5}) and considering the leading
terms in $\hbar $ we obtain the equations for the $R_{i}$ and $S_{i}$. The
results are that each $S_{i}$ satisfies the classical relativistic
Hamilton-Jacobi equation 
\begin{equation}
c^{2}\left( \partial _{j}S_{i}\right) ^{2}=\left( E-V({\bf x})\right)
^{2}-m^{2}c^{4},  \label{7}
\end{equation}
and the equation for each $R_{i}$ coming from the imaginary part is given by 
\begin{equation}
R_{i}\partial _{j}^{2}S_{i}+2\sum_{j}\partial _{j}R_{i}\partial _{j}S_{i}=0.
\label{8}
\end{equation}
Here let us define the current density 
\begin{equation}
\vec{j}_{i}=\frac{1}{m}R_{i}^{2}\nabla S_{i}.  \label{9}
\end{equation}
Eq. (\ref{8}) can be then expressed as 
\begin{equation}
\nabla \cdot \vec{j}_{i}=0.  \label{10}
\end{equation}
This equation reflects the result of the relativistic steady current. Eq. (%
\ref{7}) and Eq. (\ref{10}) are easily to solve. For the Eq. (\ref{7}), the
solution reads 
\begin{equation}
S_{i}(x_{j},E)=S_{i}(0)+\int_{x_{a}}^{x_{b}}\sum_{j}p_{i,j}dx_{j},
\label{11}
\end{equation}
where $p_{i,j}$ stands for the $j$ component of the relativistic momentum $%
p_{i}$. $p_{i}=\nabla S_{i}$ is single value on the $i$-th branch which
satisfies the relativistic Hamilton-Jacobi equation. For the Eq. (\ref{10}),
using the Gauss's law, the solution has the form 
\begin{equation}
R_{i}^{2}p_{i}d\Omega =R_{i}^{2}(0)p_{i}(0)d\Omega _{0}.  \label{13}
\end{equation}
It is obviously that $d\Omega $ is the normal cross sectional area of the
tube of trajectories. The $R_{i}^{2}(0)p(0)d\Omega _{0}$ is evaluated at
some other point on the same trajectory. This equation merely stands for the
current conservation.

Quantum mechanics required that the $\varphi ({\bf x})$ must satisfy the
single value condition. Therefore, we have for each term in Eq. (\ref{6}) 
\begin{equation}
\Delta S_{i}=h\left[ n+i\frac{\Delta \log R_{i}}{2\pi }\right] ,  \label{14}
\end{equation}
where $\Delta S_{i}$ denotes the difference between any two of its value.
This general consideration was first introduced in Keller's paper \cite{3}
for determining the quantum condition of the approximate solutions of
Schr\"{o}dinger equation.

The differences $\Delta S_{i}(x_{j},E)$ and $\Delta \log R_{i}(x_{j},E)$ are
expressible as line integrals over some closed curve in $\{x_{j}\}$ space
beginning and ending at $\{x_{j}\}.$ In term of these integrals, Eq. (\ref
{14}) becomes 
\begin{equation}
\oint \nabla S_{i}\cdot ds=h\left[ n+\frac{i}{2\pi }\oint \nabla \log
R_{i}\cdot ds\right] .  \label{15}
\end{equation}
This equation must hold for every curve in $\{x_{j}\}$ space. However, only
finite number of curves classified by the first homotopy group of space $%
\{x_{j}\}$ given the different quantum condition. At this place, we see that
the topology of configuration space $\{x_{j}\}$ determine the quantum
conditions.

Since $\nabla S_{i}\cdot ds=\sum_{j}p_{i,j}dx_{j}$, Eq. (\ref{15}) can be
rewritten as 
\begin{equation}
\oint \sum_{j}p_{i,j}dx_{j}=h\left[ n+\frac{i}{2\pi }\oint \nabla \log
R_{i}\cdot ds\right] .  \label{16}
\end{equation}
Now, let us consider the solution $S_{i}$ of the relativistic
Hamilton-Jacobi equation. The solutions in general have infinite branches.
However, most of them different each other with an constant. Only finite
different branch, say branch $i$, on which the corresponding momentum $%
\nabla S_{i}=p_{i}$ is single value. Any two branch, say branch $i$ and $j$,
are to be jointed together at all points where $\nabla S_{i}=\nabla S_{j}$.
The same considerations may be applied to the multivalue function $\log
R_{i} $ since $\nabla S_{i}=p_{i}$, and the amplitude $R_{i}\sim
p_{i}^{-1/2} $. Therefore branches on which $\nabla S_{i}$ is single value
also serves as the same branches of $\nabla \log R_{i}$. At this place, we
can omit the subscript $i$ of Eq. (\ref{16}) since each branch can be
involved in each closed integral. Also if one closed curve can be deformed
into another becomes clear in the space.

Let's return to the Eq. (\ref{13}). We see that $R^{2}$ becomes infinite
whenever $pd\Omega =0$. These points are called conjugate points. Those
conjugate points which correspond to the vanishing of $d\Omega $ are
envelope of the family of trajectories. Therefore $\nabla S$ is multivalue
near these surfaces. They are boundaries of the different branches. Those
conjugate points at which $p=0$ also form part of the boundaries of
different branch. Example for this two type conjugate points can be found in
Refs. \cite{4.1,4.2} in which each closed Kepler ellipse of the Kepler
trajectories in 3-dimensional space have three conjugate points. Two of
first order conjugate points come from the contact's place of each
trajectory with envelope and the dimensional reduction by one of the cross
section of a trajectory tube. Another second order conjugate point is the
focus of the ellipse orbit. In mathematically, the order sum of a trajectory
is called the Morse index \cite{Morse}. It is an invariant of topology.

It is well known in optics that the phase of $R$ is retarded by $\pi /2$,
i.e. $R$ is multiplied by $e^{-i\pi /2}$ on a ray, the positive direction
along a ray is the direction of $\nabla S$, which passes through a conjugate
point on which $d\Omega $ vanishes simply. Furthermore the phase is retarded
by $\pi $ on a ray passing through a focus, which is a conjugate points at
which $d\Omega $ vanishes to the second order. The similar retard phases
arise in the present discussion. The total change quantity of $\Delta \log R$
is given by $m\pi /2$. Therefore we obtain 
\begin{equation}
\frac{i}{2\pi }\Delta \log R=\frac{i}{2\pi }\oint \nabla \log R\cdot ds=%
\frac{m}{4},  \label{17}
\end{equation}
where $m$ stands for the order sum of the conjugate points of a trajectory.
When Eq. (\ref{17}) is considered, the quantum conditions of a relativistic
system become 
\begin{equation}
\sum_{j}\oint p_{j}dx_{j}=h\left[ n+\frac{m}{4}\right] .  \label{18}
\end{equation}
This result is similar to the non-relativistic results. It can be viewed as
the extension of the Keller's non-relativistic cases. This result is
important in the semiclassical solution of the relativistic quantum
mechanical problems \cite{lk}. Where the standard calculational methods of
wave mechanics always converge very slowly, yet the semiclassical
approximations are often astonishingly accurate. Furthermore, the
semiclassical method have the great merit of the describing almost all the
physics.

\section{the relativistic hydrogen atom via the relativistic quantum
conditions}

To explain the results in the above, let us consider the relativistic
hydrogen atom. Its relativistic radial momentum is given by 
\begin{equation}
\nabla S=p_{r}=\pm \frac{1}{c}\sqrt{\left[ \left( E+\frac{e^{2}}{r}\right)
^{2}-m^{2}c^{4}\right] -\left( \frac{L}{r}\right) ^{2}},  \label{19}
\end{equation}
where the $(L/r)^{2}$ is the centrifugal barrier, and $L$ is constant of
motion associated with the angular momentum. It can be decided in the
semiclassical region by comparing the asymptotic phase of $r\rightarrow
\infty $ with the phase of the free relativistic particle in the same region
and is given by 
\begin{equation}
L=l+\frac{1}{2}.  \label{20}
\end{equation}
We see that $\nabla S=p_{r}$ is real and double valued. The two branches of $%
p_{r}$ become equal at the turning points. Thus the branches for $p_{r}$
consists of two line segments joined together at their two turning points.
This space is topologically equivalent to a circle, and there is only one
primitive closed curve on it. Therefore, it has only one quantum condition.
The first order, obviously for one dimensional case, conjugate points are
given by the two turning points. We have $m=2$ for each closed orbit and by
using Eq. (\ref{18}) 
\[
I_{r}\equiv \frac{1}{2\pi }\oint p_{r}dr=\int_{r_{a}}^{r_{b}}dr\frac{1}{c}%
\sqrt{\left[ \left( E+\frac{e^{2}}{r}\right) ^{2}-m^{2}c^{4}\right] -\left( 
\frac{l+1/2}{r}\right) ^{2}} 
\]
\begin{equation}
-\int_{r_{b}}^{r_{a}}dr\frac{1}{c}\sqrt{\left[ \left( E+\frac{e^{2}}{r}%
\right) ^{2}-m^{2}c^{4}\right] -\left( \frac{l+1/2}{r}\right) ^{2}}=\left(
n_{r}+\frac{1}{2}\right) \hbar ,\quad n_{r}=0,1,2\cdots .  \label{21}
\end{equation}
Here $r_{a}\leq r\leq r_{b}$ are the turning (conjugate) points at which the
momentum change sign. The integral in Eq. (\ref{21}) can be performed. The
result is given by 
\begin{equation}
\frac{1}{\hbar c}\left[ \frac{Ee^{2}}{\sqrt{m^{2}c^{4}-E^{2}}}-\sqrt{\left[
\left( l+1/2\right) \hbar c\right] ^{2}-e^{4}}\right] =\left( n_{r}+\frac{1}{%
2}\right) .  \label{22}
\end{equation}
From this, we have the correct energy spectra of the relativistic hydrogen
atom 
\begin{equation}
E=\pm mc^{2}\left[ 1+\frac{\alpha ^{2}}{\left[ \left( n-l-1/2\right) +\sqrt{%
\left( l+1/2\right) ^{2}-\alpha ^{2}}\right] ^{2}}\right] ^{-1/2}.
\label{23}
\end{equation}
To construct $\varphi (r)$, we note the cross section $d\Omega _{0}/d\Omega
=1$ in the present case. Therefore, Eq. (\ref{13}) yields 
\begin{equation}
R=\left\{ 
\begin{array}{l}
A\frac{1}{c}\left\{ \left[ \left( E+\frac{e^{2}}{r}\right) ^{2}-m^{2}c^{4}%
\right] -\left( \frac{l+1/2}{r}\right) ^{2}\right\} ^{-1/4} \\ 
Ae^{-i\pi /2}\frac{1}{c}\left\{ \left[ \left( E+\frac{e^{2}}{r}\right)
^{2}-m^{2}c^{4}\right] -\left( \frac{l+1/2}{r}\right) ^{2}\right\} ^{-1/4}
\end{array}
\right. ,  \label{24}
\end{equation}
where $A$ is constant. The phase retardation in the above expressed by the
factor $e^{-i\pi /2}$ which occurs on passing through either conjugate
point. We finally obtain the semiclassical expression of the relativistic
Coulomb system 
\[
\varphi (r)=A\frac{1}{c}\left\{ \left[ \left( E+\frac{e^{2}}{r}\right)
^{2}-m^{2}c^{4}\right] -\left( \frac{l+1/2}{r}\right) ^{2}\right\} ^{-1/4} 
\]
\[
\times \left\{ \exp \left[ -\frac{i}{\hbar }\int_{r}^{r_{b}}dr\frac{1}{c}%
\sqrt{\left[ \left( E+\frac{e^{2}}{r}\right) ^{2}-m^{2}c^{4}\right] -\left( 
\frac{l+1/2}{r}\right) ^{2}}\right] \right. 
\]
\[
+\left. \exp \left[ -i\frac{\pi }{2}+\frac{i}{\hbar }\int_{r}^{r_{b}}dr\frac{%
1}{c}\sqrt{\left[ \left( E+\frac{e^{2}}{r}\right) ^{2}-m^{2}c^{4}\right]
-\left( \frac{l+1/2}{r}\right) ^{2}}\right] \right\} 
\]
\[
=Ae^{-i\pi /4}\frac{1}{c}\left\{ \left[ \left( E+\frac{e^{2}}{r}\right)
^{2}-m^{2}c^{4}\right] -\left( \frac{l+1/2}{r}\right) ^{2}\right\} ^{-1/4} 
\]
\begin{equation}
\times \sin \left\{ \frac{1}{\hbar }\int_{r}^{r_{b}}dr\frac{1}{c}\sqrt{\left[
\left( E+\frac{e^{2}}{r}\right) ^{2}-m^{2}c^{4}\right] -\left( \frac{l+1/2}{r%
}\right) ^{2}}+\frac{\pi }{4}\right\} .  \label{25}
\end{equation}
We may obtain the result for a relativistic particle in any one dimensional
potential by replacing the Coulomb potential $-e^{2}/r$ by $V(x)$.

\section{Concluding remarks}

In this paper we have presented the quantum conditions of the relativistic
integrable systems, starting with the approximate solution $\varphi ({\bf x}%
)=\sum_{i}R_{i}\exp \{iS_{i}/\hbar \}$ of the relativistic Klein-Gordon
equation with $R_{i}$ and $S_{i}$ being the solutions of the different
branches of the current conservation equation and the relativistic
Hamilton-Jacobian equation, respectively, and then considering the
single-valueness of the wave function $\varphi ({\bf x})$. The result can be
viewed as the extension of the non-relativistic quantum conditions of
Einstein, Brillouin, and Keller. Because of the semiclassical quantization
recently playing an important role in many fields, we hope that the result
presented in the paper will be useful in the discussions of the
semiclassical treatment of the relativistic quantum problems.

\end{document}